\documentclass[preprint,12pt]{elsarticle}
\usepackage{hyperref}
\hypersetup{
	colorlinks=true,
	urlcolor=blue,
	citecolor=blue}
\usepackage{pbox}
\usepackage[all]{xy,xypic}
\usepackage{amsfonts,amsmath,amsgen,amsopn,amsbsy,theorem,epsfig}
\usepackage{eufrak,amscd,bezier,latexsym,mathrsfs,enumerate}\usepackage[utf8]{inputenc}\usepackage[english]{babel}
\usepackage[dvipsnames]{xcolor}
\usepackage[pagewise]{lineno}
\usepackage{graphicx}
\usepackage{amssymb}

\journal{}

\title{Private key encryption and recovery in blockchain}

 \author[label1,label3]{Mehmet Aydar} 
 \address[label1]{AI Enabled Department, Huawei Turkey Research and Development Center, Istanbul, Turkey}
 \address[label5]{Softtech Reseach and Development Center, Istanbul, Turkey}
 \author[label1,label5]{Salih Cemil ÇETİN}
 \author[label2]{Serkan Ayvaz}
 \address[label2]{Department of Software Engineering, Bahcesehir University, Besiktas, Istanbul, Turkey}
 \address[label3]{Enterprise Architecture and Technology Innovation, Ford Otosan, Istanbul, Turkey \footnote{Corresponding author: maydar@kent.edu}}
 \address[label4]{Izmir Democracy University, Izmir, Turkey}
 \author[label1,label4]{Betül AYGÜN}

\def\E{\ifmmode{\mathbb E}\else{$\mathbb E$}\fi} 
\def\N{\ifmmode{\mathbb N}\else{$\mathbb N$}\fi} 
\def\R{\ifmmode{\mathbb R}\else{$\mathbb R$}\fi} 
\def\Q{\ifmmode{\mathbb Q}\else{$\mathbb Q$}\fi} 
\def\C{\ifmmode{\mathbb C}\else{$\mathbb C$}\fi} 
\def\H{\ifmmode{\mathbb H}\else{$\mathbb H$}\fi} 
\def\Z{\ifmmode{\mathbb Z}\else{$\mathbb Z$}\fi} 
\def\P{\ifmmode{\mathbb P}\else{$\mathbb P$}\fi} 
\def\T{\ifmmode{\mathbb T}\else{$\mathbb T$}\fi} 
\def\SS{\ifmmode{\mathbb S}\else{$\mathbb S$}\fi} 
\def\DD{\ifmmode{\mathbb D}\else{$\mathbb D$}\fi} 

\newcommand{\bse}{\begin{subequations}}
\newcommand{\ese}{\end{subequations}}
\newcommand{\ben}{\begin{enumerate}}
\newcommand{\een}{\end{enumerate}}
\newcommand{\bens}{\begin{enumerate*}}
\newcommand{\eens}{\end{enumerate*}}
\newcommand{\be}{\begin{equation}}
\newcommand{\ee}{\end{equation}}
\newcommand{\bea}{\begin{eqnarray}}
\newcommand{\eea}{\end{eqnarray}}
\newcommand{\baa}{\begin{eqnarray*}}
\newcommand{\eaa}{\end{eqnarray*}}
\newcommand{\bc}{\begin{center}}
\newcommand{\ec}{\end{center}}

\theoremstyle{corollary}

\theoremstyle{lemma}

\theoremstyle{proposition}

\theoremstyle{axiom}

\theoremstyle{conjecture}

\theoremstyle{example}

\theoremstyle{definition}

\theoremstyle{remark}

\begin{document}
\begin{frontmatter}

\begin{abstract}
The disruptive technology of blockchain can deliver secure solutions without the need for a central authority. 
In blockchain protocols, assets that belong to a participant are controlled through the private key of an asymmetric key pair that is owned by the participant. 
Although, this lets blockchain network participants to have sovereignty on their assets, 
it comes with the responsibility of managing their own keys. 
Currently, there exists two major bottlenecks in managing keys; $a)$ users don't have an efficient and secure way to 
store their keys, $b)$ no efficient recovery mechanism exists in case the keys are lost. 
In this study, we propose secure methods to efficiently store and recover keys. 
For the first, we introduce an efficient encryption mechanism to securely encrypt and decrypt the private key using the 
owner's biometric signature. For the later, we introduce an efficient recovery mechanism using biometrics and 
secret sharing scheme. By applying the proposed key encryption and recovery mechanism, asset owners are able to securely 
store their keys on their devices and recover the keys in case they are lost. 

\end{abstract}

\end{frontmatter}

\section{Introduction}
\label{Int}

In a blockchain network, trust is embedded in the network itself. Therefore, blockchain reduces the cost of ``trust'' by eliminating the third parties traditionally needed for providing trust.
This is achieved through the cryptographic linking structure of the blocks, distribution of the ledger and a consensus algorithm.
Many initiatives exist aiming to replace centralized solutions with blockchain based distributed solutions. 
As a result, when centralized authorities are removed as the provider of ``trust'', individuals have more sovereignty on their assets while cost associated with trust is reduced. 
However this imposes more responsibility on the network participants on managing their own keys. 

Asymmetric keys play a vital role in identifying network participants and controlling the assets in a blockchain network. 
An asymmetric key pair consists of a public key which can be shared with anyone and a corresponding private key which must be stored hidden.
In blockchain protocols, an asymmetric key pair is assigned to a network participant, and participants are identified by the public key of the asymmetric key pair, while asset ownerships and transfers are managed through self-controlling of the private key. 

Despite the ever-growing adoption of the blockchain technology, major problems persist in storing and recovering private keys that have a negative impact on usability of the blockchain technology, and the security of assets in the network. Traditional private key storing mechanism includes key memorization, cold storage, keeping the key digitally in plain form, keeping the key remotely through a wallet provider, and keeping the key in a symmetrically encrypted digital wallet. Memorization is challenging as the private keys are too long for humans to memorize. For instance, in Bitcoin system, private keys' length are 256 bits in hexadecimal which can be represented in 64 characters in the range 0-9 or A-F. Cold storage is as secure as the physical material where the private key is stored on, and has inefficiencies in terms of usability as the key retrieval is challenging from the cold storage. While digital key keeping in plain form has more usability, it is the least secure option as the digital devices could be open and susceptible to hacking and security breaches.

Wallets are usually responsible for the process of private and public key pair creation.
In web wallet services, private keys that belong to the clients are encrypted and stored on related servers. Web wallets allow users to control their assets from any web browser or mobile platforms.
Despite the usability advantages, storing the key remotely through a digital wallet provider is a centralized solution, and is only as secure as the remote party, which is trusted for keeping the private keys safe. 
In desktop wallets like Electrum \cite{turuani2016automated}, each user locally keep their private key with the encryption option.
Regular symmetrically encrypted digital wallets provide better security. However, the user must remember the password used in the encryption for the key retrieval. If forgotten, it would be impossible to retrieve the private key. This mechanism does not consider biometrics of the key owner. 

As a matter of the fact, it is crucial to move away from traditional key storage mechanisms towards a more user-friendly and secure key storage approach, which incorporates the biometrics of the key owner along with a distributed key recovery mechanism. This paper focuses on secure and user-friendly storage of private keys, and private key recovery methods. 
The main contributions of this study include the following:
\begin{itemize}
	\item  We provide a framework for secure encryption and decryption of private keys using biometric fingerprints
	\item  We propose a biometric-based distributed private key recovery mechanism in blockchain.
	\item We review existing solutions in this domain, and described problems persists in traditional private key storage and recovery mechanisms.
\end{itemize}

The rest of the paper is organized as follows.
Section \ref{sec:Blockchain} briefly describes blockchain technology, specifically concentrating the usage of keys and describes what key owners' control by securely managing their keys, and what is compromised if the keys are lost. In section \ref{sec:Methodology}, we describe our solution. In section \ref{sec:RelatedWork}, we review the existing work in the domain, and we follow by conclusion.

\section{Blockchain Overview} 	\label{sec:Blockchain}
World met blockchain with bitcoin which is popular for its proven solid functionality of decentralized peer-to-peer digital asset transfer \cite{nakamoto2008bitcoin}.  
Blockchain protocol gets its form with blocks which are  
chained with hashes. This chain of blocks structure provides tamper-proofness and it doesn't permit any changes on historical records. On the other hand, each block consist of transactions and some unique information about the block. In this section, we emphasize the key points of blockchain protocols.

Transactions are valuable data transfers in blockchain protocols. Depending on the protocol, transactions may contain various type of data including but not limited to financial value, health data, a log record or identity information.
Blocks are bundle of transactions with some block specific information such as the number of the block, previous block's hash, transactions' merkle root \cite{merkle1980protocols}, a timestamp and a nonce value.

Chained blocks include cryptographic chain mechanism which uses hash functions. Every block has a hash output of previous block in the ledger. Hash functions are deterministic and one way functions which always generate the same output for the same input. The function outputs are unique for an input and do not contain any meaningful information about the input. For each different input, hash functions generate a completely different output. 




Hash outputs can be assumed as abstract or fingerprint of a piece of information. In the chain, each block header contains the hash value of the previous block, constituting a chain of blocks. This mechanism provides immutability of the data stored in the ledger. When data in a block changes, the block hash also changes. Consequently, the next block's hash output also change in the chain. If any data changes in block $a$, the chain is broken after block $a$.


\subsection{Public and Private Keys in Blockchain} 
In blockchain protocols, public keys are commonly used as address, account number, id etc. Therefore, naturally it can be shared with other users in the ecosystem. Private keys are used for signing transactions by its owner \cite{loera2015method}. 
In blockchain applications, generally elliptic curve digital signatures (ECDSA)  \cite{johnson2001elliptic} or similar algorithms are used to create public and private key pairs. 
In common, a participant's digital value ownership means the values which can be digitally signed and used as input in transactions by the participant. 

\begin{figure}
	\includegraphics[scale=0.8]{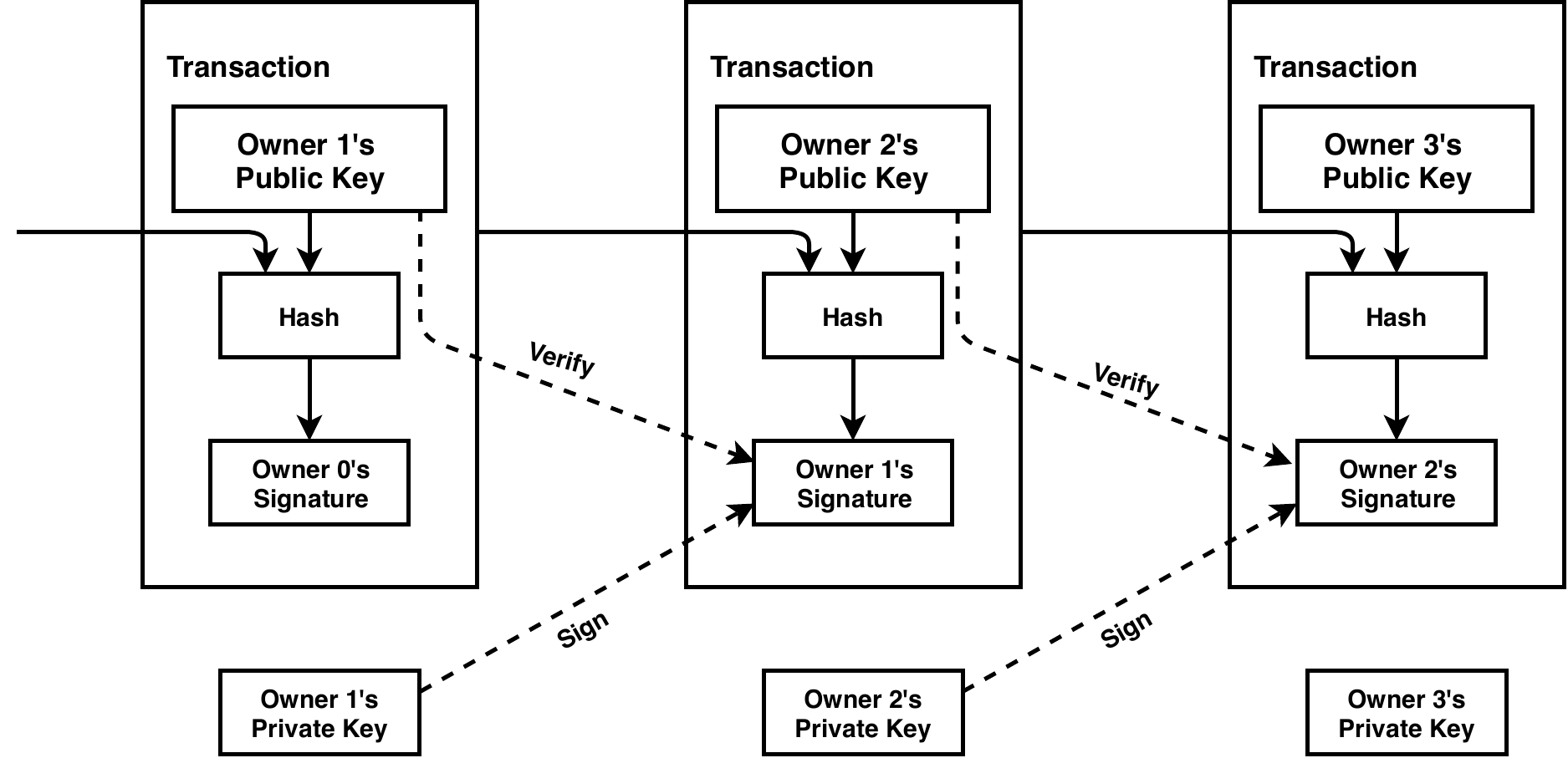}
	\caption{Bitcoin transaction scheme \cite{nakamoto2008bitcoin}}
	\label{fig:Bitcoin_transaction_scheme}
\end{figure}

As it’s described in Satoshi’s bitcoin white paper \cite{nakamoto2008bitcoin}, when a client sends a coin to another one, the owner actually does not send any asset to anyone. Instead of sending a digital coin, the wallet reassigns an amount of coin and disseminates the transaction to the network so that it can create and assign new coins to the receiver. In order to reassign the coins, each transaction is signed by transaction owner's private key and is verified using owner's public key. Figure \ref{fig:Bitcoin_transaction_scheme} shows transaction verifications in bitcoin transaction scheme. 
 
To provide a true mint based model system, transactions are declared publicly and explicitly in public-permissionless blockchains like bitcoin. 

\subsubsection{What is Managed with Private Key} 	\label{sec:ManagedKeys} 
In financial applications such as bitcoin or ripple, users sign their transactions with private keys. Each coin (asset) in a transaction has a public key on it. Since private key is associated with public key, private key holder is the owner of the coin. Similarly in health care projects, people manage their sensitive health information. Distributed digital identity projects manage sensitive identity information. In supply chain projects, participants manage critical tracking records, and in real estate projects ownership of real estate properties is managed. 

Since various kind of assets are bound up with private keys in blockchain projects, the safety of keys is vital. In case of key compromise, the attacker is able to spend money or reach/share/sell sensitive information, or create fraudulent records, or own and take the initiative of private properties. The possibility of compromise scenarios and their potential consequences require blockchain developers to build more secure protocols and key preservation strategies. 

\subsection{Transactions in Blockchain}
Transactions are basic units or atomic events of blockchain protocols. Blockchain protocols usually have their own type of assets, which are transferred through transactions. As an example in bitcoin system, transactions include coin transfers, while in sovrin \cite{SovrinWhitePaper_whatgoesonledger}, verifiable credentials and identity management information are processed through transactions. 
Since transactions are atomic events of blockchain applications, the ownership of transaction is critically important. In each transaction, depending on protocol's transaction architecture, there is one or more addresses as related to user endpoint. These addresses are generally public keys of users. Public key or its derivatives are used as address or endpoint. In blockchain applications, personal information is never used to provide anonymity. As an example, there must be at least two public keys in a normal bitcoin transaction to manifest the transaction which is processed between two users. 

Since there is no central authority, each user is responsible for creation of their own transactions. A transaction is firstly created by the owner within validation rules. This transaction is then checked by protocol's authorized nodes and is processed if validated. 

\begin{figure}
	\includegraphics[scale=0.8]{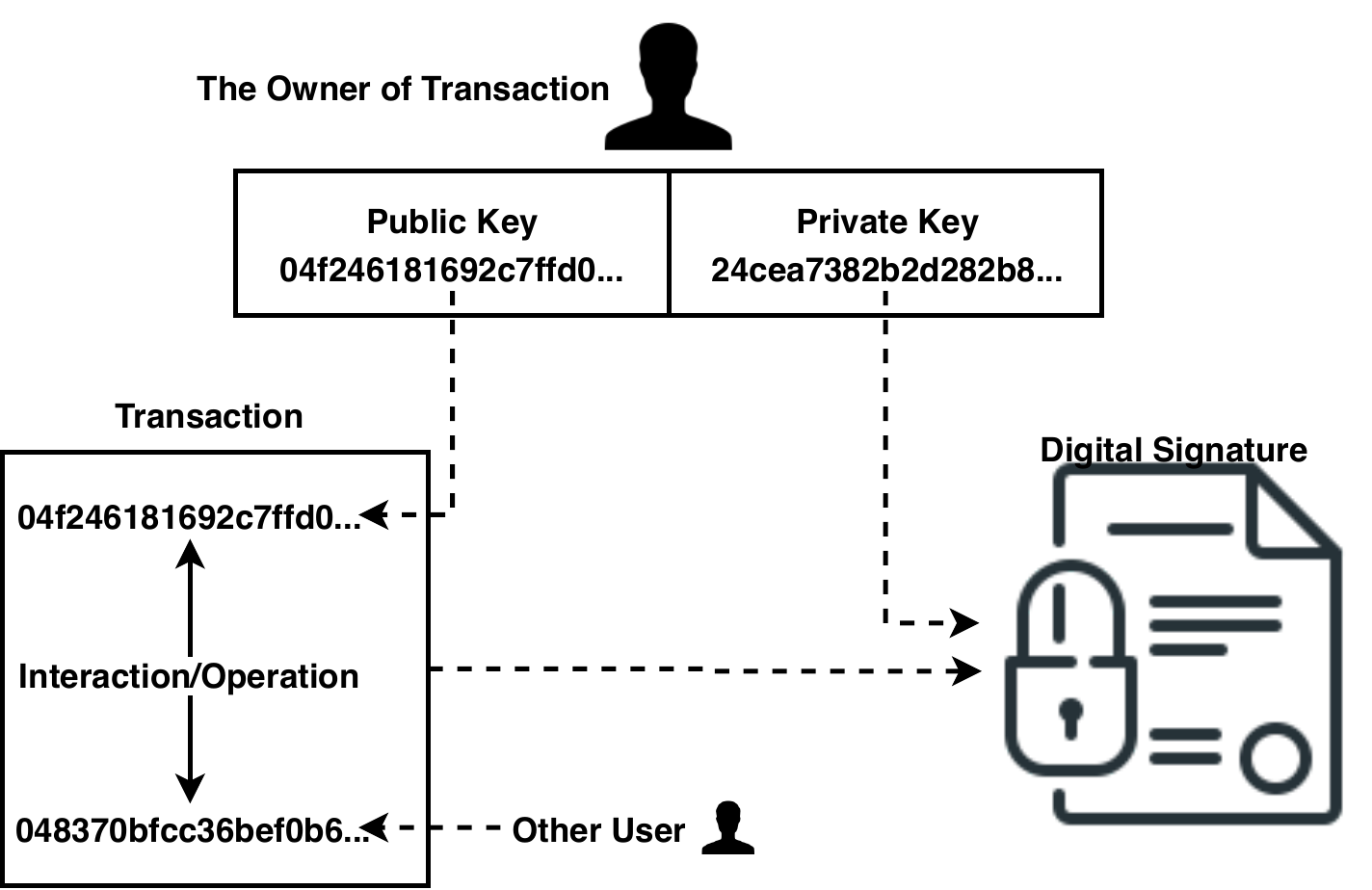}
	\caption{Simplified scheme of a blockchain transaction}
	\label{fig:blockchain_transaction}
\end{figure}

In Figure \ref{fig:blockchain_transaction}, a basic transaction between two users is shown. As demonstrated in the figure, the transaction is between 04f246181692c7ffd0... and 048370bfcc36bef0b6... addresses. Each address represents a real world user without revealing any personal information. Therefore, transactions are generally transparent in blockchain protocols. When a transaction is generated, it must also be digitally signed by its owner. 

Digital signatures are mathematical techniques that verify authenticity and ensure integrity of digital contents \cite{rivest1978method}. A valid digital signature shows that the content is original as sent, and the sender is known. These features basically points out authentication and integrity issues. 
Digital signatures provide two benefits in blockchain protocols: everyone is able to verify transaction sender and sender is not able to deny the transaction. Transactions transparency only can be handled with user anonymity and digital signatures.

\section{Methodology}	\label{sec:Methodology}

In order to verify data authenticity or integrity, a user only needs digitally signed data with a signature and the signer's public key. From this point of view, a transaction owner can just sign the transaction using its private key. Then, other nodes in the network are able to verify the owner and integrity of transaction by using just the public key of the sender of transaction. Thus, public keys can be shared explicitly between users and private keys should be kept hidden and safe. 

In general, there are three different approaches for keeping security of holding private keys for users. In the first approach, adding additional security layer to reach the private keys stored on the device. Biometric authentication is used to open a private key. In the second approach, stored private keys are also encrypted with the biometric data. Instead of encrypting machine holding security keys, the encryption of the private keys is performed. In the last approach, private keys are generated by implementing biometric data into known prominent cryptography algorithms including DES, RSA. In this study, we use second approach for private key encryption, and we utilize a distributed key recovery mechanism for private key recovery. The authentication is done using biometric data based on features of fingerprint, which is a popular authentication technique.
    
\subsection{Encryption and Decryption of Private Keys Using Fingerprint} 
\label{encryption-decryption-private-key}

\begin{figure}
	\includegraphics[scale=0.7]{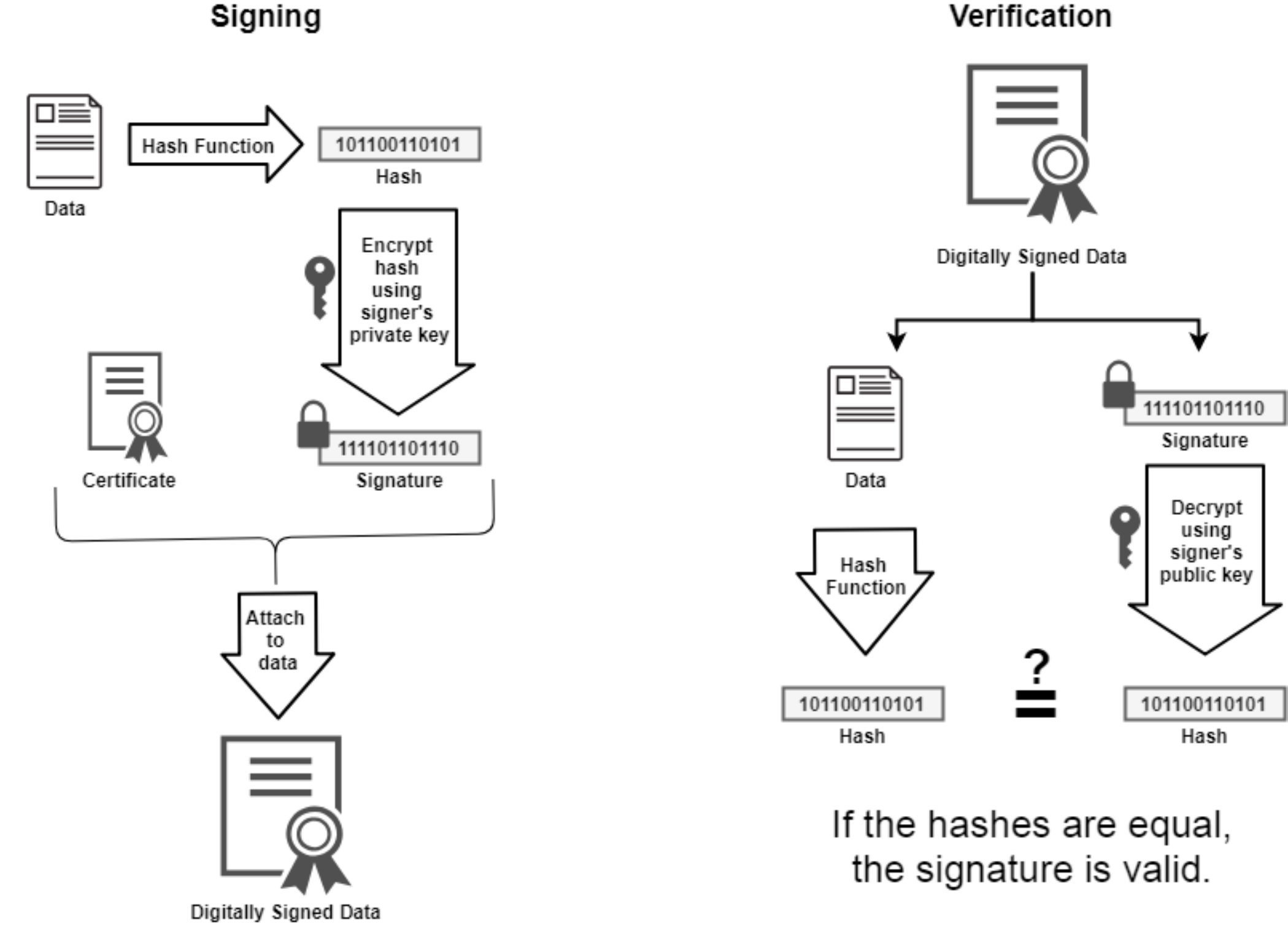}
	\caption{Digital signature signing and verification}
	\label{fig:signature_verification}
\end{figure}

In blockchain protocols, assets must be related with owner's public key which represents the owner digitally and anonymously in DLT environment. Transactions must be digitally signed before it is published to the network. As shown in Figure \ref{fig:signature_verification}, a private key is sufficient for a user to digitally sign a document. 

Symmetric encryption and decryption is straightforward using conventional symmetric encryption methods such as Data Encryption Standard (DES), in which encryption and decryption is done using the same key.   
In our approach, we use symmetric encryption to encrypt and decrypt private keys, and we automatically generate the key used in symmetric encryption using owner's fingerprint. 
Fingerprints, as a biometric trait, is unique and offers usability advantages over traditionally selected pass codes. 
However, certain concerns regarding privacy, security, and applicability have to be dealt with when using fingerprints. 

In fingerprint recognition systems, there exists two main phases: registration and matching. 
Registration step includes registering the original fingerprint image, while matching step includes matching the candidate fingerprint image against the registered image. In both phases, the fingerprint image is preprocessed, transformed, and hashed. Since it is probabilistically hard for two fingerprints taken at different times to have the same hash value (even if they match), an efficient error correction mechanism is utilized. 

\subsubsection{Preprocessing}
\label{preprocessing}
Preprocessing includes image enhancement (filtering, binarization and thinning), minutiae points extraction, core points detection, and minutiae alignment according to the core points.
Purpose of enhancement step is to compensate for scratches and noises, 
and produce a binary fingerprint image to accurately detect its structure. 
We apply Gabor filter \cite{finger_enhancement_hong1998fingerprint} method, 
in which each pixel is filtered according to estimated ridge frequency and ridge orientation. 
Enhancement step is proceeded with binarization using a threshold variable, and thinned that fixes the ridge lines width to one pixel. 

The minutiae detection algorithm traverses the enhanced image to detect whether a pixel represents a minutiae by checking its surrounding 8-neighboring pixels. 
If the pixel is on a ridge and has 1 neighboring ridge pixel then the pixel represents a ridge ending type of minutiae, 
on the other hand, if the pixel is on a ridge and has 3 neighboring ridge pixel then the pixel represents a bifurcation type of minutiae. 

Core points' position and orientation are needed in order to reliably align the minutiae points with respect to these points as reference.  
The core points (poincare index) of a fingerprint are special pixels that represents the centers. 
Loop, delta and whorl are types of core points. 
We use fingerprint core detection method suggested by Kawagoe et al. \cite{kawagoe1984fingerprint}, 
which divides the image into sub-regions, obtains direction patterns and computes core points over a closed curve. 
For a pixel$(x,y)$, it sums the difference between adjacent local ridge orientation angles in its 8-neighborhood. 
Based on the result of the calculation with a small threshold: 
\begin{itemize}
\item $(x,y)$ is not a core point if result is $0$, \\
\item  $(x,y)$ represents a whorl type core point if result is $2\pi$, \\
\item  $(x,y)$ represents a loop type core point if result is $\pi$, and \\
\item  $(x,y)$ represents a delta type core point if result is $-\pi$. 
\end{itemize}

In the minutiae alignment step, each minutiae point is rotated using the rotation of axes in two dimensions. A minutiae point $(x,y)$ is rotated counterclockwise with respect to a core point $(cx,cy)$ with an orientation angle $\theta$ using the matrix multiplication as below:
$$
\quad
\begin{pmatrix} 
x' \\
y' 
\end{pmatrix}
=
\quad
\begin{pmatrix} 
cos\theta & sin\theta \\
-sin\theta & cos\theta 
\end{pmatrix}
\quad
\begin{pmatrix} 
x-cx \\
y-cy
\end{pmatrix}
$$

\subsubsection{Cartesian Transformation}
\label{cartesian-transformation}
An efficient implementation of a biometric system needs to be revocable, since revocability is a must have feature for password systems for privacy and security purposes. 
Fingerprints as a biometric signature are permanently associated with the owner, and if stolen, all systems previously used with the fingerprint signature are in danger.  
Therefore, we apply cartesian transformation to transform the minutiae points using a one-way, irreversible function to make the fingerprint system cancellable. 
Instead of storing the original fingerprint image, we store the transformed version along with the transformation parameters. 

In cartesian block transformation, the 2D coordinate system on which the minutiae points are represented is divided into blocks of regular size. 
Initially, minutiae points are placed in the blocks based on their locality, such that closer minutiae points are placed in the same or neighboring blocks. 
Later, the transformation is achieved by shuffling the blocks using matrix multiplication, and arranging minutiae points based on the new block locations. 

\begin{figure}
	\includegraphics[scale=1]{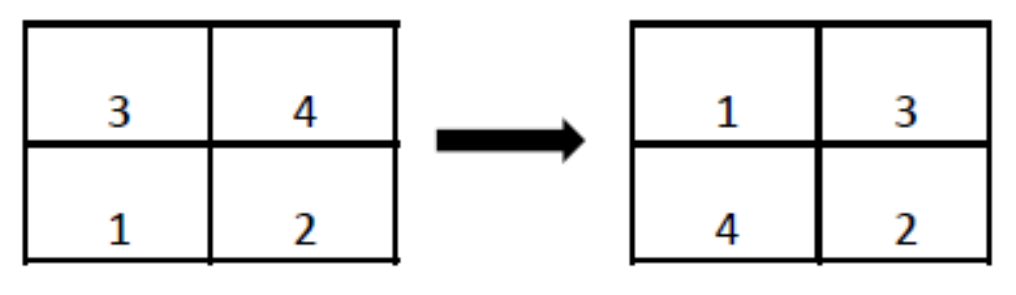}
	\caption{Cartesian blocks transformation}
	\label{fig:cartesian-transformation}
\end{figure}

In our implementation, 2D coordinate system is divided into a $H x W$ size of blocks.  
Initial cartesian blocks are numbered from $1$ to $|H x W|$ which is represented by a matrix $C$ of size $1 x |HxW|$, and a transformation matrix $M$ of size $|HxW| x |HxW|$ is randomly generated having values of either $0$ or $1$. As an example, let $H = 2$ and $W = 2$, then $C = [1, 2, 3, 4]$. Then the matrix multiplication with the randomly generated matrix of $M$ is shown below:

$$
\quad
C'
=
\quad
\begin{pmatrix} 
1 & 2 & 3 & 4
\end{pmatrix}
\quad
\begin{pmatrix} 
0 & 0 & 0 & 0 \\
0 & 1 & 0 & 0 \\
1 & 0 & 0 & 1 \\
0 & 0 & 1 & 0
\end{pmatrix}
\quad 
=
\begin{pmatrix} 
3 & 2 & 4 & 3
\end{pmatrix}
$$
which means the minutiae points previously placed in cartesian block $1$ are mapped to $3$ and $2$ are mapped to $2$ again. Similarly, points in cartesian block $3$ are mapped to $4$, and $4$ are mapped to $3$ in the transformed space, as demonstrated in figure \ref{fig:cartesian-transformation}. It is also possible for multiple cartesian blocks to be mapped to the same cartesian block in the transformed space. Cartesian blocks are numbered per their locations in the 2D coordinate system.

In the registration phase, instead of saving minutiae points' original locations, their transformed locations are saved along with the transformation parameters. The transformation parameters include the boundaries of the original fingerprint image and the transformation matrix. It is important to note that, the original cartesian block for a given minutiae point is not saved during the registration phase. However, during matching, for candidate fingerprint template, minutiae points' original cartesian blocks are kept to be utilized in recovery process of the reed-solomon error erasure coding. 

\subsubsection{Reed-solomon Error Correction}
\label{reed-solomon}
\begin{figure}
	\includegraphics[scale=1]{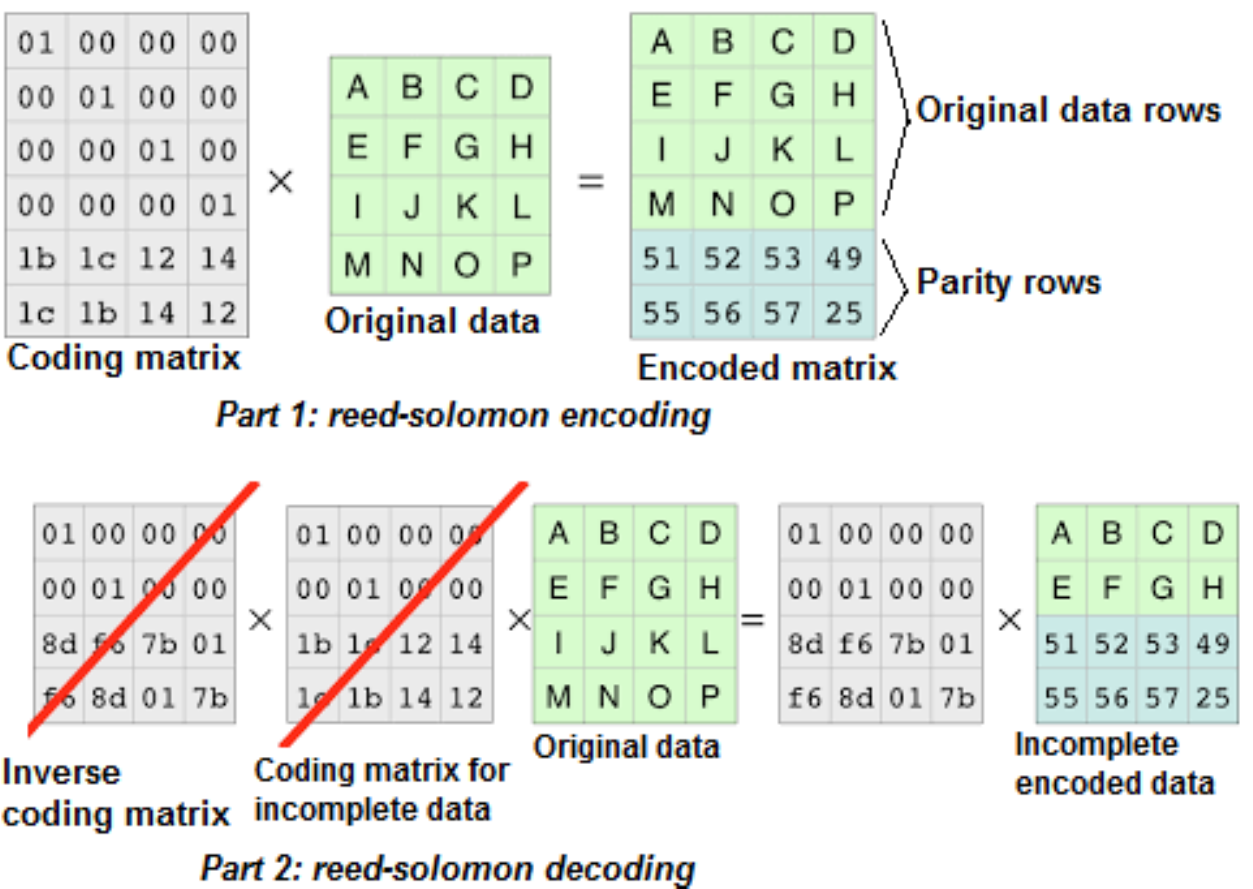}
	\caption{Example of reed-solomon encoding and decoding (adapted from \cite{beach2015backblaze}.)}
	\label{fig:reed-solomon-example-encode-decode}
\end{figure}

Reed-solomon \cite{reed1960polynomial} is an error correction (erasure coding) mechanism. For a given input, it produces parity data, in a way that it can reproduce original input even if some parts are missing. Many modern storage systems, such as Linux RAID and Facebook's cold-storage utilize reed-solomon. Reed-solomon breaks the message into $n$ equal pieces and constructs an input matrix, where $n$ is the height of the matrix. Then, it generates a coding matrix of size $n+k$, $k$ is being the number parity rows. First $n$ rows of coding matrix has $1$s in the diagonal and $0$s for the rest of the matrix cells. The coded data is created by multiplying the coding matrix with the original matrix. Because of the diagonal $1$s in the coding matrix, the first $n$ rows of the coded data is the same as the original message, and the last $k$ rows are parity. Thus, one row of the coding matrix generates a corresponding row of original data. Therefore, when some rows in the original message is missing, the corresponding rows in the coding matrix and the coded matrix are removed, and the matrix multiplication equation with the original data on the left side still remains valid. Later, inverse matrix of the new coding matrix is generated, and multiplied with the each side of the new equation. In the end, the original data matrix is produced on the left side of the equation. Figure \ref{fig:reed-solomon-example-encode-decode} depicts an example of reed-solomon encoding of a given input data of ``ABCDEFGHIJKLMNOP'', and the reed-solomon decoding when ``IJKLMNOP'' is missing from the input data.

In our implementation, the hashes of the minutiae points are the input data. We perform reed-solomon mechanism per each of the pre-transformed cartesian blocks, having the hashes of the minutiae points as input data inside the blocks. 
By using the hashes of the minutiae points original minutiae points of original template are never revealed, a mechanism which preserves the privacy of the fingerprint owner. 
We also perform an overall reed-solomon implementation for all of the pre-transformed rectangular, having the resulting hash of each cartesian block as input data. This way, the missing hashes of the minutiae points can be recovered for each cartisian block. Consequently, we are able to calculate an overall hash of the fingerprint system that we utilize in the matching process. Overall hash is the symmetric key to be utilized in the symmetric encryption of owner's private key. 
Our goal is to regenerate the same key for the same person during the matching phase. 

\subsubsection{Matching}
In the matching phase, we follow a number of steps to determine whether a given candidate fingerprint image produces the same overall hash value as with the original fingerprint image. The candidate fingerprint image goes through the same preprocessing and transformation steps as the original fingerprint image, as described in sections \ref{preprocessing} and \ref{cartesian-transformation}. In the cartesian transformation, the same transformation parameters (boundaries and transformation matrix) are used as in the registration of the original fingerprint image. Moreover, in contrast to the registration phase, the original pre-transformed cartesian block numbers are kept for the candidate fingerprint image. 

The matching algorithm compares the transformed minutiae points of candidate fingerprint template with the transformed minutiae points of original fingerprint template. The comparison is done separately for each of the cartesian blocks. Geometrically closer minutiae points would be transformed to the same cartesian block in both original and candidate fingerprint template. Therefore, the minutiae points in cartesian block number $x$ in the transformed candidate fingerprint template are only compared with the minutiae points in cartesian block number $x$ in the transformed original fingerprint template. The comparison is done using the equality check of the minutiae point types, and the euclidean distance with a reasonable threshold. If a match found, the original cartesian block number of the minutiae point that belongs to the candidate fingerprint template is used to reverse the transformation of the minutiae point that belong to the original fingerprint template. In this way, the original minutiae locations are recovered for the matched minutiae points. 

The recovered minutiae points for each of the cartesian blocks are gone into the reed-solomon decoding process as explained in section \ref{reed-solomon}, and a resulting hash is generated. If the generated hash is the same as the hash generated in the registration phase, then the fingerprint images match. Using this hash value and the same symmetric algorithm used in the encryption, encrypted private key is decrypted. Implementation code described in our methodology is available for research purposes \footnote{http://bit.ly/cancellable-fingerprint-encryption}.

\subsection{Private Key Recovery} \label{sec:Recovery}
Cryptography in blockchain protocols is heavily based on public and private key pairs. 
Since public key is open to public, key recovery in distributed ledger technology generally implies private key recovery. Private key theft and losses are major security problems in blockchain systems. In other words, compromise of private key leads to losing ownership of the assets associated with the private key. Therefore, the recovery of private keys is of utmost importance in blockchain based systems. To provide an efficient, secure and scalable key recovery in blockchain, we propose a distributed key recovery mechanism for encrypted private keys based on Shamir's Secret Sharing (SSS) Scheme \cite{shamir1979share}. 

\begin{figure}
	\includegraphics[scale=0.4]{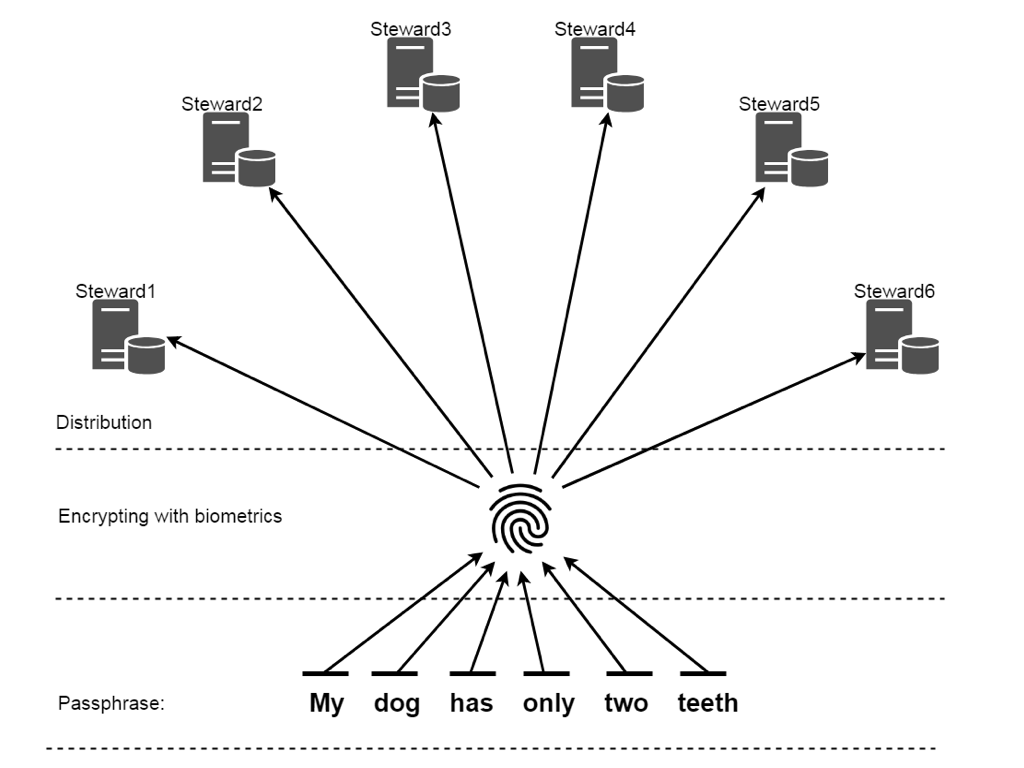}
	\caption{Distributed Key Recovery Example}
	\label{fig:DistributedRecovery}
\end{figure}


The distributed key recovery mechanism of our approach is illustrated using an example in Figure \ref{fig:DistributedRecovery}. In the figure, a passphrase is encrypted using biometrics. The encrypted text is divided into 6 pieces and distributed to 6 parties called stewards. Thus, to recover the original passphrase, the pieces must be collected from stewards and it should be decrypted using biometrics. 

In this approach, biometrics provides an extra layer of security. However, it requires for all stewards to be available in order to retrieve the original passphrase. In order to tackle this problem, we use Shamir's Secret Sharing Schema which add more flexibility for recovery.

\subsubsection{Recovery using Shamir's Secret Sharing Scheme}
According to SSS Scheme, data $D$ is divided into $n$ pieces, and $k$ pieces of $D$ can reconstruct $D$, but even $k-1$ pieces reveals no information about $D$ \cite{shamir1979share}. 
By using this method, owners can divide their secret into $n$ pieces and distribute them to $n$ different location. Even if some pieces are lost, any $k$ pieces will be sufficient to recover the secret. 
SSS can be applied to recover a private key. In distributed digital identity systems, there are actors named as stewards who are trusted nodes in the network. Stewards' services can be used as distributed pieces' locations. 

\begin{figure}
	\includegraphics[scale=0.6]{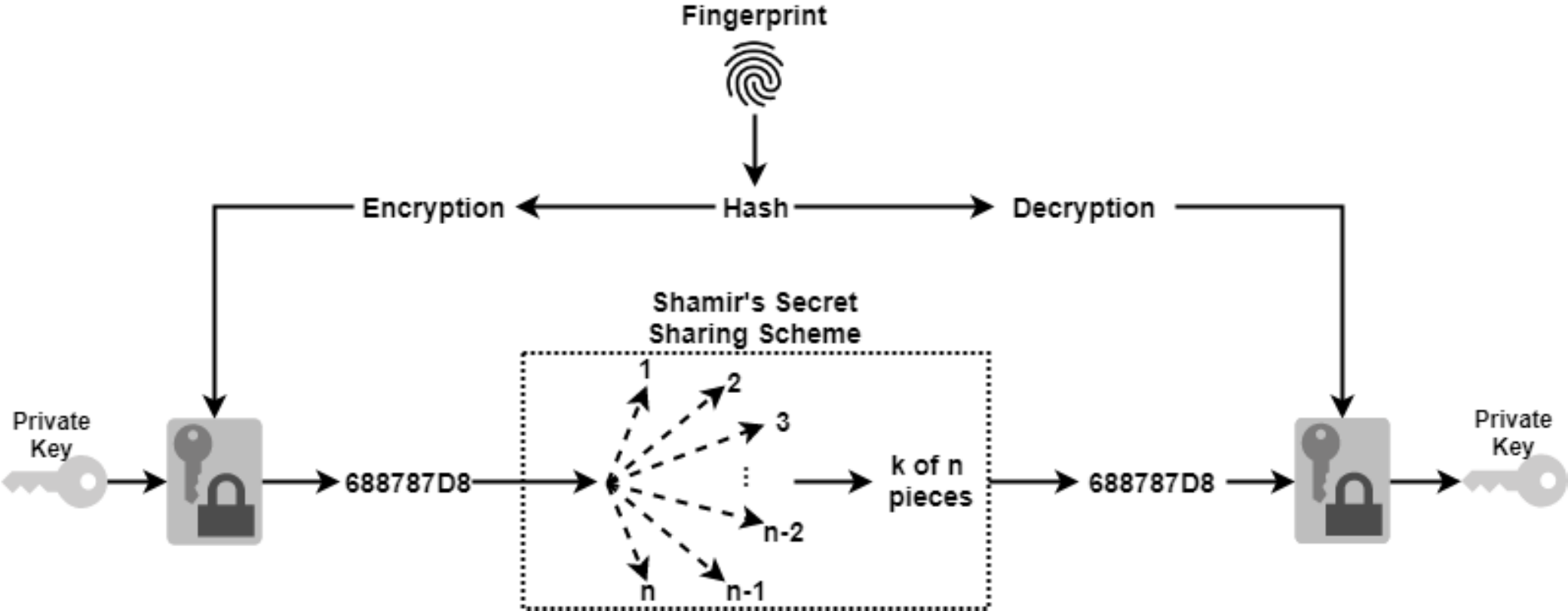}
	\caption{Proposed Key Recovery Mechanism}
	\label{fig:KeyRecovery}
\end{figure}

In our approach, biometric information is used to protect a private key by encrypting it. We first create a symmetric key utilizing fingerprint data of the owner, and we encrypt the private key with the symmetric key as detailed in section \ref{encryption-decryption-private-key}. After the encryption process, the encrypted output is split into $n$ pieces, and each of the pieces is distributed to different and secure locations such as steward services. For recovery, any $k$ parts of $n$ encrypted pieces are sufficient to recover the private key. Once the encrypted private key is recovered, the same symmetric key is regenerated using fingerprint and used to decrypt the encrypted private key. In Figure \ref{fig:KeyRecovery}, the overview of our proposed private key recovery mechanism are illustrated in details.  

Our key recovery is based on two main principles: encryption with biometrics and distribution. Since each encrypted biometric hash and private key pair is distributed to a set of stewards for secure storage, even when a steward node in the blockchain network is intruded, the data stored in the node would not be sufficient to decrypt private key. Because the attackers would not know the locations of other stewards to complete the entire hash key. Furthermore, despite the possibility of having accessed to all steward nodes containing the combined hash value, the private key could not be decrypted without user’s biometric information which is kept in user’s local device. In other words, there are two separate factors of security.
 

\section{Related Work} \label{sec:RelatedWork}
Selecting appropriate biometric data to create key pairs is another issue that must be considered. Researchers have investigated several biometric features in biometric-based cryptographic key generations \cite{jagadeesan2010secured}.
 
There are very few studies that integrate biometric traits into RSA keys. In the study, Je-Gyeong proposed a method for generating keys of digital signature (public and private key) from biometric. 
Some others investigated iris texture as a biometric feature for generating cryptographic key. Rathgeb and Andreas proposed an approach using bits of the iris code for deriving biometric cryptographic key \cite{rathgeb2011context}.
Janbandhu et al. derives signature keys from the code generated by using the 512 byte iris biometric data invented by J. Daugman \cite{janbandhu2001novel}. Similarly in another study, Boyen et al. also investigated the iris texture as biometric trait \cite{boyen2005secure}. In the study by Sarkar et al., biometric authentication was used for obtaining asymmetric cryptography keys \cite{7791193}.

Monrose et al. proposed a method using users' voice as biometric trait \cite{monrose2001cryptographic}. Their system regenerates the key from the user’s voice by asking the user to repeat the same pass phrase. In the study by Chen and Chandran, the image of user’s face was used in biometric key generation \cite{chen2007biometric}. The same face image is required for regeneration of the key in the future.

From a different perspective, Perera et al. offered a new technique that combines digital signature with public key cryptography \cite{perera2017biometric}. This new technique was implemented for RSA and ECC algorithms. In another study, the proposed algorithm was developed utilizing inner productions computation with error correction mechanism \cite{lan2009biometrics}.

Mjaaland et al. suggested an approach in which public keys are extracted from users' fingerprints \cite{gligoroskibiocryptics}. Another fingerprint that belong to the same user is processed to generate the private keys. The method is resilient to the variations in the samples to generate the same resulting key.  

Trotter proposed a fingerprint matching approach utilizing cartesian block transformation with reed-solomon erasure coding \cite{trotter2007mapping}. In the study, reed-solomon algorithm is performed on the entire original fingerprint template. Also, in the reed-solomon decoding process exact locations of the original minutiae points are recovered in contrast to our system, in which we only recover the hash of the original minutiae points.

In a related study, Kwon et al. also proposed a digital signature based on biometric data without holding them in hardware devices \cite{kwon2004practical}. Studies that are done on biometric creation were underway many years due to the difficulties in achieving the uniformity of the biometric data from the noise. In the study \cite{bansal2018enhanced}, the authors process biometric image first to provide the uniformity of the unstable biometric traits. In biometric cryptosystems, images of the biometric traits are taken and preprocessed. Then, using the preprocessed image, minutiae of the biometric are extracted. Later, the image minutiae points are transformed into a 1024 prime number generator to generate 2048 cryptographic key used in RSA chipper algorithm.
 
Kayva et al. claimed that if the face recognition system for biometric affirmation is considered, then AES gives more sublime security than RSA and DES \cite{kavya2018survey}. To avoid problems from occurring due to the certification authorities, identity based public key cryptography and certificate-less PKI was also proposed \cite{liu2018biometrics}. The communication phase between two peers has two phases. In the first phase (initialization phase), users produce public keys from biometric data. In the second phase (authentication and key agreement), they authenticates identities. Due to the nature of the blockchain, there are no authorities to keep the certification. 

\begin{figure}
	\includegraphics[scale=0.7]{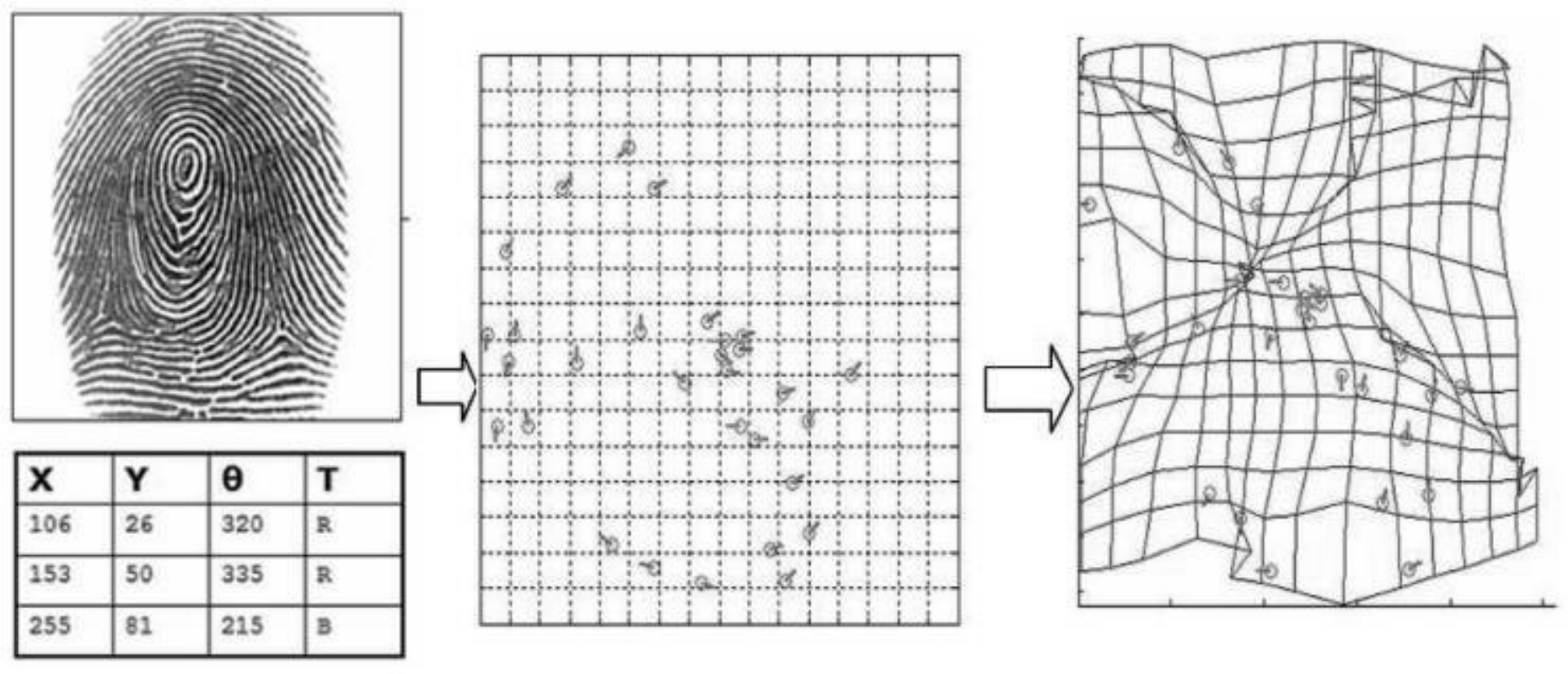}
	\caption{Fingerprint image surface folding transformation \cite{ratha2001enhancing}}
	\label{fig:Fingerprint_image_surface_folding_transformation}
\end{figure}

Security and privacy are major concerns in biometric-based cryptography. Due to irrevocable nature of biometric traits, these systems must provide revocability \cite{soutar1999biometric}. 
As biometric data are inherent, they cannot be changed if compromised. Thus, in order to satisfy revocability of a generated key, biometric data must not be directly associated with the biometric properties. As shown in Figure \ref{fig:Fingerprint_image_surface_folding_transformation}, Ratha et al. proposed a fingerprint image surface folding transformation approach that extract minutiae positions from fingerprint image and generates cancelable biometric templates. Since biometric templates are transformed, even when the data compromised, the original biometric data cannot cross-matched with biometric databases. Similarly, Barman et al. offered an approach using session-based biometric keys, meaning that another unique key should be generated in a new session using the same biometric data \cite{barman2015fingerprint}.

In some studies, researchers explored applying more than one biometric traits instead of using only one biometric trait. Jagadeesan et al. proposed multimodal biometric system that generates a 256-bit secure cryptographic key using a combination of features from iris texture and minutiae points from the user’s finger prints \cite{jagadeesan2010secured}. In the study of Manjunath et al., they propose multimodal approach of  biometric. For instance they use iris and fingerprint, speech and signature, face and voice etc. In the study, iris and fingerprint modalities are used and evaluated under FAR, FRR and accuracy \cite{manjunath2018analysis}. Also, the study conducted by Yik-Herng proposes multi modal biometric systems that combine iris and fingerprint with IFO hash fusion method \cite{khoo2018multimodal}.
Iris trait is unique for each individual even for identical twins. Also, false acceptance rate (FAR), the rate of invalid matches, is lower than all other biometric traits like fingerprint and face. Voice trait is a composite of both behavioral and physical biometrics. Behavioral part differentiates in time due to the factors like medical conditions and age. In contrast to token or password-based systems, biometric matching does not work well every time due to the false matching or false mismatching.

Bhattacharyya et al. provided a review on biometric authentication technologies \cite{bhattacharyya2009biometric}. They found that fingerprint based systems had 2\% FAR and 2\% false reject rate (FRR). On the other hand, face recognition system resulted in 1\% FAR and 10\% FRR. According to their study, iris technologies achieved the best accuracy score in both FAR and FRR with 0.94\% and 0.99\%, respectively. Similarly, when comparing the biometric traits, Deborah et al. stated that iris recognition is the most suitable for mobile locking followed by fingerprint and face biometric data \cite{hui2018assessment}. The results of the study of Yik-Herng et al.showed that although the proposed method yields better results than unimodal fingerprint biometric system, it does not perform as well as iris recognition system \cite{manjunath2018analysis}. However, the proposed method contributes to the security aspect. 

The study by Naser et al. also stated that indexing structure of iris surpass indexing structure of fingerprint. According to their study, the hit rate was improved from 97.0\% up to 99.8\% in multi-modal approach, and 98.3\% respectively for fingerprint and indexing \cite{damer2018fingerprint}. Based on the previous research, it appears that the iris biometric system performs slightly better in terms of accuracy when compared to the other biometric systems, in which the fingerprint comes closest the most. Since asymmetric encryption (public key cryptography) is a newer and more secure technique than symmetric encryption methods, asymmetric encryption techniques are used in our study.
 
Smart card based biometric user authentication schemes have also been proposed \cite{clancy2003secure}.
The biometric data and keys are stored in a smart card for regeneration of keys in the future. However,
smart card based approaches have portability issues as carrying physical card is an additional
burden. Also if compromised, they pose security threats for biometric data.
Fingerprint technology provides very accurate results \cite{ratha2001enhancing}. Also, Jain et al. claimed that no biometric data is better than the other traits because all have own strengths and weaknesses, and performance of biometric data selection related with the type of application \cite{jain2006biometrics}. However, the matching accuracy of the fingerprint has been shown to be very high \cite{maltoni2009handbook}.

\section{Conclusion}
Through blockchain implementations, the hegemony of central authorities is reduced. While this is positive for reducing the cost of providing ``trust'' in the system, it increases the responsibility of the network participants on managing their keys. In blockchain, valuable data are locked to the public key of the owner, and can only be unlocked for spending with associated private key (asymmetric cryptography.) 
In this study, we focused on laying a foundation for securely encrypting and decrypting private keys used in controlling asset ownership in the blockchain using a symmetric key generated from owner's fingerprint, and a distributed private key recovery system utilizing secret sharing scheme supported by biometric. 
We reviewed existing solutions in this domain, and described problems persists in traditional private key storage and recovery mechanisms in terms of security, usability and privacy. Our methodology includes the concepts of revocable fingerprints and erasure codes for key encryption, and distributed secret sharing scheme for key recovery. 
As for future work, we aim to integrate the proposed solution on mobile applications with white-box cryptography.





\bibliographystyle{apalike}
\bibliography{sample}

\end{document}